\documentclass[a4paper,12pt]{article}
\usepackage{amsmath}
\usepackage{amsfonts}
\usepackage{amssymb}
\usepackage{latexsym}
\usepackage{epsfig}
\usepackage{graphicx}
\usepackage{oldgerm}
\usepackage{theorem}
\usepackage{bm}

\def\N{\mathbb{N}}

\def\bra{\langle}
\def\ket{\rangle}

\def\PS{\mathrm{P}_{\rm S}}

\theorembodyfont{\itshape}

\newcommand{\SSC}[1]{\section{#1}\setcounter{equation}{0}}

\begin{document}

\title{
\bf 
Extinction and Metastability of
Pheromone-Roads in Stochastic Models
for Foraging Walks of Ants
}
\author{
Saori Morimoto
\footnote{
Department of Physics,
Faculty of Science and Engineering,
Chuo University, 
Kasuga, Bunkyo-ku, Tokyo 112-8551, Japan
}, \, 
Makoto Katori
\footnote{
Department of Physics,
Faculty of Science and Engineering,
Chuo University, 
Kasuga, Bunkyo-ku, Tokyo 112-8551, Japan;
e-mail: 
makoto.katori.mathphys@gmail.com
}, \,
Hiraku Nishimori
\footnote{
Meiji Institute for Advanced Study of Mathematical Sciences, 
Meiji University, 
Nakano, Nakano-ku, Tokyo 164-8525, Japan;
e-mail: nishimor2@meiji.ac.jp
}
}

\date{14 February 2025}
\pagestyle{plain}
\maketitle

\begin{abstract}
Macroscopic changes of group behavior of eusocial insects
are studied from the viewpoint of non-equilibrium 
phase transitions.
Recent combined study of experiments and mathematical
modeling by the group led by the third author
suggests that a species of garden ant 
switches the individual foraging walk from
pheromone-mediated to visual-cues-mediated 
depending on situation.
If an initial pheromone-road between the nest 
and food sources is a detour, 
ants using visual cues can pioneer shorter paths.
These shorter paths are reinforced by pheromone
secreted by following ants, and then the detour 
ceases to exist.
Once the old pheromone-road extincts, 
there will be almost no chance to reconstruct it.
Hence the extinction of pheromone-road
is expected to be regarded as 
a phase transition to an absorbing state.
We propose a discrete-time model on 
a square lattice consisting of
switching random walks 
interacting though time-dependent pheromone field.
The numerical study shows that
the critical phenomena of 
the present extinction transitions of pheromone-roads
do not seem to belong to the
directed percolation universality class
associated with the usual absorbing-state transition.
The new aspects are cased by the coexistence
and competition with newly creating pheromone-roads.
In a regime in the extinction phase,
the annihilating road shows metastability and
takes long time-period to be replaced by a new road. 

\vskip 0.2cm

\noindent{Keywords:} 
Extinction transitions of roads; 
Switching interacting particle systems; 
Non-equilibrium phase transitions;
Metastability of annihilating roads

\end{abstract}


\SSC
{Introduction}
\label{sec:introduction}

The group behaviors of so-called \textit{eusocial insects}
(\textit{e.g.} ants, bees, and termites) have been
extensively studied not only in biology \cite{HW90}
but also in statistical physics as 
cooperative phenomena in systems consisting of
permanently moving and mutually 
interacting units \cite{Nis15}. 
Persistent motion is a common feature of
living systems, but recently several models composing of 
self-propelled units have been studied
to simulate physical and chemical systems.
Such \textit{interacting self-propelled particle systems}
provide many interesting phenomena 
in their collective motions,
which can not be realized as the thermal or
chemical fluctuation phenomena in equilibrium systems.
In non-equilibrium statistical physics, 
study of self-propelled particles
in both living and non-living worlds
is one of the most challenging 
research subjects \cite{MKN25,NSN17,VZ12}.

Ants in colonies share various types of information
through direct and indirect communication
and exhibit highly organized group behaviors
and perform complex tasks \cite{HW90}.
Foraging is one of the most intensively studied subjects
in such interesting group 
behaviors \cite{EMTKN24,Nis15,VZ12}.
Various species of ants establish lengthy foraging
trails through a positive feedback process
caused by the secretion and tracking of recruit pheromone.
Although such `pheromone-roads' enable
them to efficiently shuttle
between the nest and food sources, 
a recent combined study of experiments
and mathematical modeling for a species of
garden ant reported by the research group of
the third author of the present paper 
(the Nishimori group) 
suggested a situation-dependent switch of the primarily
relied cues from recruit pheromone to visual cues.
The latter refer to landmarks, 
the polarization angle of the sun or moon,
textures of the edges of crowded plants or woods, 
and other visual stimuli \cite{Nis15}.
The Nishimori group used \textit{Lasius Japonicus},
a species of garden ant found in Japan. 
From another experiment, it is considered that
this species of ant can determine landmarks 
in the order of 10 cm \cite{Shi13}.
Plastic boxes measuring
$23 \times 12 \times 3.5$ cm were prepared,
which are equipped with 
special floors and walls covered in plaster
to maintain internal humidity .
Approximately 200 to 300 ants were accommodated 
in each box, which was shielded by a black plate 
to shield the ants from light.
Using column chromatography, 
recruit pheromone was extracted
from the remaining ants collected from the
same natural colony.  
In the box, a food source was placed at a separate 
position from the nest.
A conflicted situation was made so that the
homing direction for ants from the food source
indicated by recruit pheromone is different from
that by visual cues as explained in the following.

\begin{figure}[ht]
\begin{center}
\includegraphics[width=0.6\textwidth]
{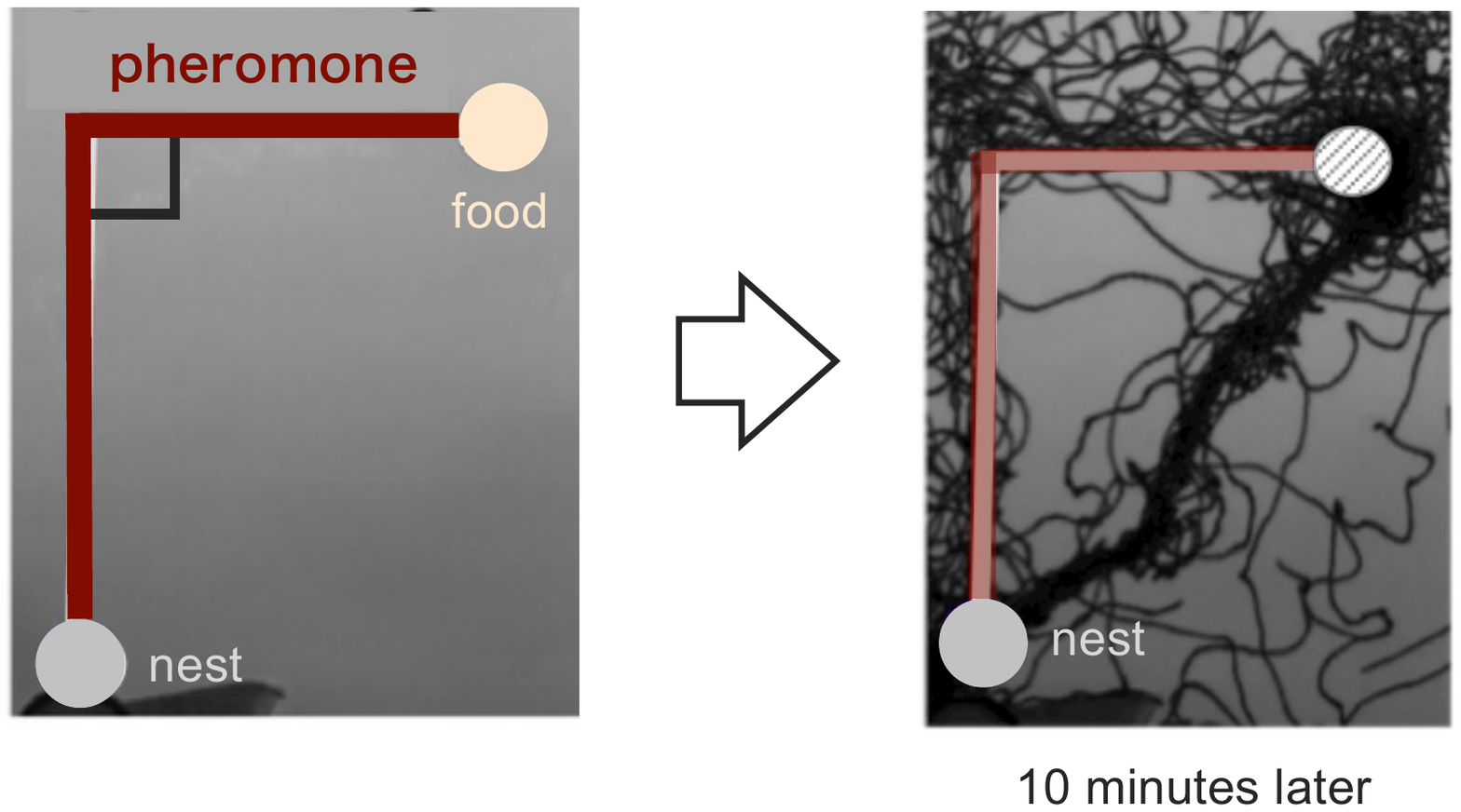}
\end{center}
\caption{Experimental setting is shown in the
left picture, where the initial pheromone-road 
was prepared as a detour 
connecting the nest and the food source 
as indicated by red lines.
The trajectories of ants observed during the middle
period (approximately 10 minutes) of an experiment
are shown in the right picture \cite{Nis15}.}
\label{fig:ant_paths}
\end{figure}

\begin{description}
\item{1.} \,
At the beginning of experiment,
the preliminarily extracted recruit pheromone
was applied along a line
connecting the nest and the food source
with one folding point with turning angle 
$90^{\circ}$. 
We call this $\Gamma$-shaped line
a initially prepared \textit{pheromone-road}.
See the left picture in Fig.\ref{fig:ant_paths}. 

\item{2.} \quad
Ants in the present setup could
recognize the direction of the nest from the food source.
This direction along the optimal path from the
food source to nest 
(i.e., the diagonal direction) is different from
the direction indicated by the pheromone-road
(i.e., the horizontal direction).
\end{description}

The motions of foraging ants were recorded by 
a video camera. 
For further details on the experiments, 
refer to the original paper \cite{Nis15}.
The results are shown in 
the right picture of Fig.~\ref{fig:ant_paths}.
The trajectories of foraging ants
recorded during the middle period 
(approximately 10 minutes) 
of each experiment with a duration of 60 min. 
A direct path in the diagonal direction 
was established between the nest and food source.

In a previous paper \cite{EMTKN24}, 
a discrete-time stochastic model was proposed
to simulate selection processes of foraging paths by ants.
This model was 
inspired by the concept of switching interacting 
particle systems recently introduced by \cite{Flo22}
and have been extensively studied 
for population dynamics
\cite{dH22a,dH22b,Gre22,Gre23,Len12}.
We consider an interacting random walk model
of hopping particles on a square lattice 
\cite{CD98,HSIMT18,MD99,MOSS92,NMI97,Tai88}.
The exploring ants are expressed by
random walkers in exploring period
drifted in the direction from the nest to the food source, 
while ants returning from the food source to the nest
are expressed by 
random walkers in homing period
drifted in the opposite direction.
We regard the exploring and homing ants
following the pheromone-mediated walk
as random walkers in the \textit{slow mode}, and 
the homing ants using the 
visual-cues-mediated walk as in the 
\textit{fast mode} in the notion of 
switching particle systems \cite{Flo22};
\begin{align}
\mbox{pheromone-mediated walk} 
&\iff \mbox{slow mode}, 
\nonumber\\
\mbox{visual-cues-mediated walk}
&\iff \mbox{fast mode}.
\label{eq:correspond}
\end{align}
During the exploring period,
all particles are in slow mode, whereas
when they arrive at the food source
and return to the nest, 
a part of particles switch their mode from
slow to fast. 
These fast-mode particles only realized in
the homing period represent
``pioneering ants'', who can find shorter paths
to the nest. 

Once shorter paths connecting
the food source and the nest are established
by pioneering ants,
these paths are reinforced by recruit pheromone
secreted by followers of ants
both in the exploring and homing periods, and then
a new pheromone-road is established. 
This implies that the old pheromone-road,
which was a detour but artificially prepared 
in the beginning of the experiments
will be annihilated after a sufficiently long time-period.
In the present paper, we study such 
\textit{extinction phenomena of road} 
as \textit{non-equilibrium phase transitions
into absorbing states} \cite{CD98,Hin00,MD99}.
The non-equilibrium critical phenomena
associated with the transitions
from fluctuating phase into absorbing states
has been extensively studied, since they 
form a wide universality class for variety of
models including directed percolation models,
epidemic models (e.g., contact processes),
forest fire models, heterogeneous catalysis models
and so on (see \cite{Hin00} and references
therein). 
The experimental evidence of the absorbing phase 
transitions has been also reported in
turbulent liquid crystals 
\cite{LSAJAH16,ST16,TKCS07,TKCS09}.

In the present paper, we study 
time evolution of the pheromone-road
using the switching random walk model on 
a lattice introduced in \cite{EMTKN24},
but here we simplify the switching mechanism.
In both studies, all walkers are slow particles
during the exploring period walking from the nest
to the food source. 
In the previous study \cite{EMTKN24},
after getting foods
the ants in the homing period 
switch from slow particle to fast particle
with probability $\gamma_{\rm sf}$ and
they can switch back 
from fast to slow with probability $1-\gamma_{\rm sf}$,
where $\gamma_{\rm sf} \in [0, 1]$.
The switching between two modes
can be repeated during the homing period. 
On the other hand, in the present study,
when each slow particle in the exploring period
arrives at the food source, it remains in
the slow mode also in the homing walk
with probability $p \in [0, 1]$ and
switches to be in the fast mode with probability $1-p$,
where $p \in [0, 1]$.
Once each ant have chosen the mode of slow or fast,
it is fixed during homing period.
This simplification has allowed
us to simulate longer-term behavior of 
the model. 
In the previous study, we simulated the
behavior of random walkers expressing individual ants,
which are mutually correlated though the 
time-dependent pheromone field.
In the present study, we are interested in 
the pheromone field on the 
initially prepared pheromone-road
and newly constructing roads, 
which evolves in time relatively slowly 
effected by rapid motions of ants.
In the present picture, collection of ants 
will be regarded as 
a transmission medium which mediates the
interaction among pheromone fields
at different points in each road and
between separated roads. 

When $p$ is small, large portion of
particles arriving at the food sources
are changed to be in the fast mode. 
Since the fast particles do not follow the initially prepared
pheromone-road, the pheromone-road
will extinct, and a new pheromone-road
is being constructed by particles.
Once the old pheromone-road extincts, 
there will be almost no chance to reconstruct it,
since the new road connecting the food source and
the nest is reinforced by pheromone secreted
by ants (all particles in the exploring period
and slow particles in the homing period)
trailing the new road and hence
the old detour will not be followed by any ant (particles).
Hence the extinction of pheromone-road
is expected to be show a transition to an absorbing state
studied in non-equilibrium statistical mechanics
\cite{CD98,Hin00,MD99}.
The critical phenomena
are then expected to be 
in the universality class called
the \textit{directed percolation} (DP) 
\textit{universality class} \cite{Hin00}. 
Let $\rho(t, x)$ be the local intensity 
at position $x$ on the
initially prepared pheromone-road at time $t >0$,
which is assumed to be normalized as
$\rho(t, x) \in [0, 1]$.
If the above mentioned expectation is valid, 
$\rho(t, x)$ will obey the
following partial differential equation
in the mean-field theory
\cite{CD98,Hin00,MD99}, 
\begin{equation}
\frac{\partial}{\partial t} \rho(t,x)
=-\mu \rho(t, x) 
+ p \rho(t, x)(1-\rho(t, x)) + D \nabla^2 \rho(t, x),
\label{eq:MF1}
\end{equation}
where $\mu$ is a constant rate of evaporation of
pheromone and $D$ is the diffusion constant.
We can see that the first two terms in the
right-hand side of \eqref{eq:MF1} is written as
\begin{equation}
- p \left[
\rho(t, x)-\frac{p-\mu}{p} \right] \rho(t, x),
\label{eq:MF2}
\end{equation}
which implies that the critical value is given by
$p_{\rm c}=\mu$ 
and for $p > p_{\rm c}$,
the uniform stationary solution
$\rho=(p-\mu)/\mu$ is obtained
in the mean-field approximation.
The transition is continuous and 
$\rho$ is considered as an order parameter
with the critical exponent
$\beta^{\rm DP-MF}=1$.
We assume that the system will be
invariant under time change $t \to \kappa x$
associated with the dilatation, the
deviation from criticality $p-p_{\rm c}$, 
and the local density $\rho$, 
\begin{equation}
t \to \kappa t, \quad
x \to \kappa^{\nu_{\perp}/\nu_{\parallel}} x,
\quad (p-p_{\rm c}) 
\to \kappa^{-1/{\nu_{\parallel}}} (p-p_{\rm c}) ,
\quad \rho \to \kappa^{-\beta/\nu_{\parallel}} \rho.
\label{eq:scaling1}
\end{equation}
It is easy to verify that \eqref{eq:MF1} is 
invariant under the rescaling \eqref{eq:scaling1} 
with the critical exponents
\begin{equation}
\beta^{\rm DP-MF}=1, \quad
\nu_{\parallel}^{\rm DP-MF}=1, \quad
\nu_{\perp}^{\rm DP-MF}=1/2.
\label{eq:MF3}
\end{equation}

In the present paper, we report the numerical study
of our model in which the finite-size scaling 
associated with the invariance of the system
under \eqref{eq:scaling1} is applied
to evaluate the critical value of parameter $p$
and critical exponent.
We have estimated a nontrivial critical value
\begin{equation}
p_{\rm c} \fallingdotseq 0.750, 
\label{eq:pc}
\end{equation}
and the critical exponents,
\begin{equation}
\beta \fallingdotseq 0.49, \quad
\nu_{\parallel} \fallingdotseq 0.30.
\quad \nu_{\perp} =0.
\label{eq:exponents}
\end{equation}
This result implies that even though 
the extinction of pheromone-road can be
regarded as a non-equilibrium phase transition
to an absorbing state, the critical phenomena
do not belong to the DP universality class
studied so far.
The deviation from the DP universality class will be
due to the coexistence of the initially prepared 
pheromone-road and a newly constructed pheromone-road
which connecting the food source
and the nest by shorter paths.

The paper is organized as follows.
In Section \ref{sec:model} we explain
our discrete-time interacting random walk model
on a square lattice for foraging ants.
There the switching parameter $p \in [0, 1]$
is precisely defined.
The numerical analysis of our simulations
is reported in Section \ref{sec:annihilation},
where the extinction phenomena of initially prepared 
pheromone-road are studied as
the non-equilibrium phase transition
into an absorbing state.
The finite-size scaling is applied to evaluate
the critical values $p_{\rm c}$ and critical
exponents.
The effect from creation of new pheromone-road
to the extinction transition of
old road is discussed in Section {sec:creation},
where four types of observation-time dependence
of survival probability of initially prepared pheromone-road
are considered 
depending on the four regimes of parameter $p$. 
Concluding remarks and future problems
are given in Section \ref{sec:future}.

\SSC
{Model on a Square Lattice}
\label{sec:model}
\subsection{Switching random walks interacting through pheromone field }
\label{sec:algorithm}
\subsubsection{General setting} 

\begin{figure}[ht]
\begin{center}
\includegraphics[width=0.4\textwidth]{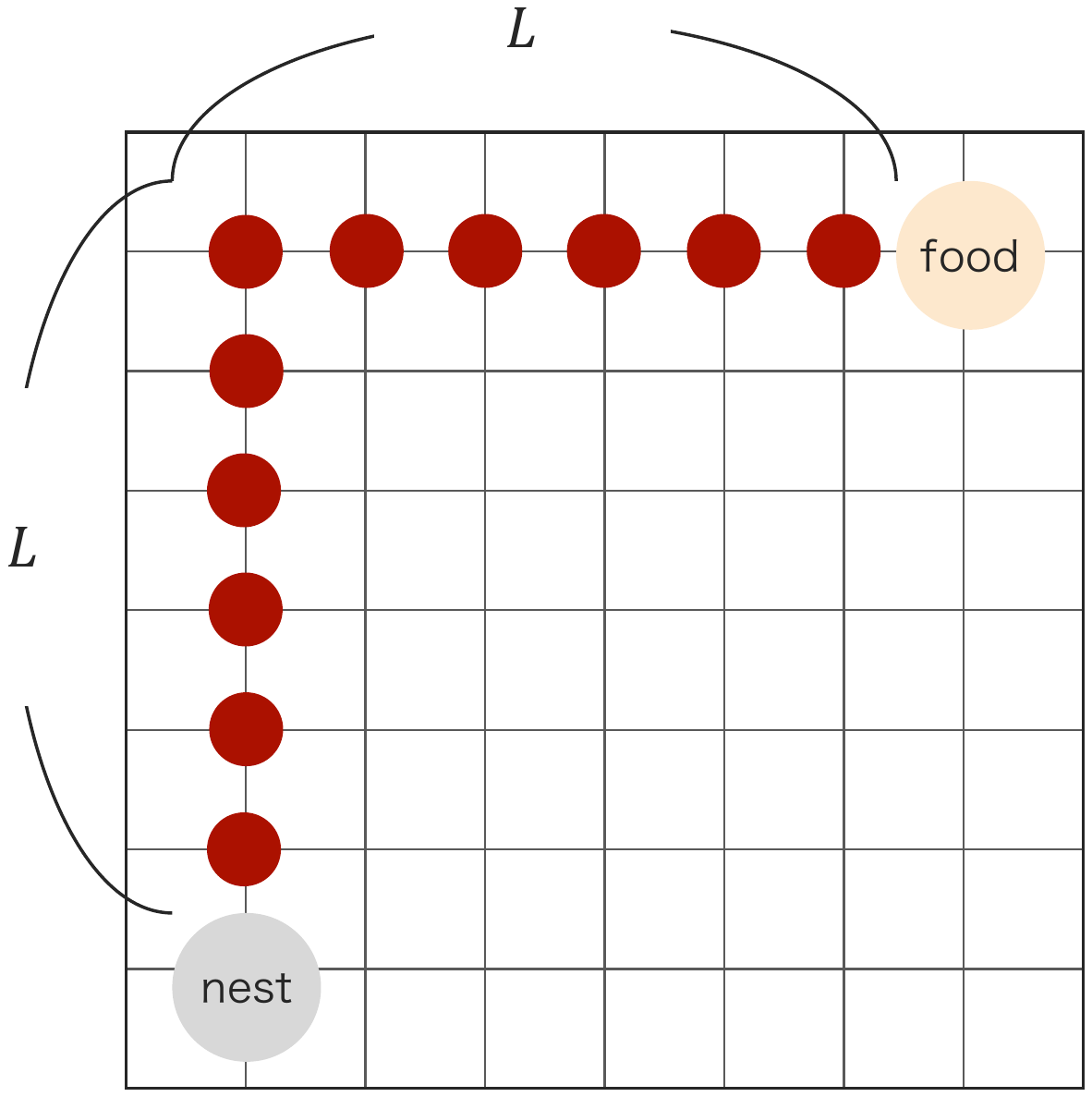}
\end{center}
\caption{We consider the $L \times L$ square lattice
and make the values of pheromone field $f$ 
have the large value $f_0$ at the vertices
only in the $\Gamma$-region at $t=0$.}
\label{fig:lattice}
\end{figure}

Consider the set of vertices in the
$L \times L$ square region 
on a square lattice, 
\begin{equation}
V_L:=\{v=(x, y): x \in \{0, 1, \dots, L \},
y \in \{0, 1, \dots, L \} \},
\label{eq:sites}
\end{equation}
where the total number of vertices is
$\sharp V_L=(L+1)^2$. 
The Euclidean distance between two vertices
$v_1, v_2 \in V_L$ is written as $|v_1-v_2|$.
On $V_L$, 
we consider a set of drifted random walks 
of a given number of particles 
$N \in \N:=\{1, 2, \dots\}$, 
which are interacting with each other through 
a time-dependent pheromone field.
Any exclusive interaction between particles are,
however, not imposed, and hence 
each vertex $v \in V_L$ can be occupied by
plural particles at the same time.

We assume that
the nest and food source are 
located at the origin $v_{\rm n}:=(0, 0)$ 
and at the most upper-right vertex 
$v_{\rm f}:=(L, L)$, respectively; 
see Fig.~\ref{fig:lattice}. 
$N$ particles start from $v_{\rm n}$ and perform
drifted walk toward $v_{\rm f}$. 
After arriving at $v_{\rm f}$, each particle immediately
starts drifted walk from $v_{\rm f}$ toward $v_{\rm n}$
and then come back to $v_{\rm n}$. 
In this way, 
particles shuttle between $v_{\rm n}$ and 
$v_{\rm f}$ for a given time period $T$.
When the particle is in a time period
for walking from $v_{\rm n}$ to $v_{\rm f}$,
we say that it is in an exploring-period,
and when it is in a time period
for walking from $v_{\rm f}$ to $v_{\rm n}$,
it is said to be in a homing-period.

We introduce the two types of hopping rules
of drifted random walk 
representing the slow mode and 
the fast mode, respectively. 
The particles are labeled by $j=1, \dots, N$.
At each time $t \in \N_0:=\{0,1,\dots\}$, 
the state of a particle is specified by 
the location: $v \in V_L$,
and the types of hopping : 
in the slow mode (s) or fast mode (f).
Hence, at each time $t \in \N_0$, 
the $N$-particle configuration 
is given by a set 
of $N$ pairs of random variables:
\begin{equation}
X_j(t)=(v_j(t), \sigma_j(t)),
\quad
v_j(t) \in V_L, \quad
\sigma_j(t) \in \{{\rm s}, {\rm f}\},
\quad j=1,2, \dots, N.
\label{eq:X2}
\end{equation}
We define a discrete-time stochastic process
$X(t)=\{X_j(t): j=1,2, \dots, N\}$, $t=1,2,\dots, T$,
as follows.

\subsubsection{Time-dependent pheromone field}

The pheromone is put on vertices.
Its intensity on each vertex $v \in V_L$ at time $t$ 
is expressed by $f(v, t)$, which we call
the pheromone field. 
Define the $\Gamma$-shaped subset of $V_L$,
\begin{equation}
\Gamma_{L}
= \{ (0, y) \in V_L: 0 \leq y \leq L \}
\cup
\{(x, L) \in V_L: 0 \leq x \leq L \}.
\label{eq:Gamma1}
\end{equation}
We call $\Gamma_{L}$ 
the $\Gamma$-region.
We introduce parameters,
$f_0, f_1, f_2$, and $\delta t$. 
The initial state of the pheromone field is given by
\begin{equation}
f(v, 0)=
\begin{cases}
f_0, &
\mbox{if $v \in \Gamma_{L}$},
\cr
0, &
\mbox{if $v \in V_L \setminus \Gamma_{L}$},
\end{cases}
\label{eq:Gamma3}
\end{equation}
where $V_L \setminus \Gamma_{L}$ denotes
the complement of $\Gamma_{L}$ in $V_L$.
Hence, the $\Gamma$-region represents
the originally prepared `pheromone-road';
see Fig.~\ref{fig:lattice}. 

In the exploring-period, ants do not secrete any
pheromone, while 
in the homing-period, 
each ant secretes $f_1$ pheromone once
in $\delta t$ steps.
That is, at each vertex $v \in V_L$, 
secretion of pheromone by an ant
is represented by the following increment of the intensity,
\begin{equation}
f \to 
f+f_1, \quad \mbox{once in $\delta t$ 
steps in a homing-period}.
\label{eq:phe1}
\end{equation}
The pheromone are evaporating; 
at each time step $t \to t+1$, 
\begin{equation}
f \to f- f_2.
\label{eq:evaporation}
\end{equation}

\subsubsection{Hopping rule in the exploring-period} 
When ants are in the exploring-period,
they are all in the slow mode; 
$\sigma_j(t) \equiv {\rm s}$. 
For each ant, $X_j(t)=(v_j(t), {\rm s})$, we define
a set of the nearest-neighboring vertices of $v_j(t)$ as
\begin{align}
\Omega^{\rm s}_j(t) &:=
\begin{cases}
\{v \in V_L: |v-v_j(t)|=1\}
=\{(1,0), (0,1)\},
&
\mbox{if $v_j(t)=v_{\rm n}$},
\cr
\{v \in V_L: |v-v_j(t)|=1, v \not= v_j(t-1)\},
&
\mbox{if $v_j(t) \in V_L \setminus \{v_{\rm n}\}$}.
\end{cases}
\label{eq:Lambda_s}
\end{align}
At each time $t$, we list out the intensities of pheromone
at all vertices in $\Omega^{\rm s}_j(t)$ and add positive
fluctuations to them,
\begin{equation}
f(w, t) \to f(w, t) + \delta f,
\quad w \in \Omega^{\rm s}_j(t).
\label{eq:fluctuation}
\end{equation}
Here $\delta f$ are independently and identically 
distributed following the $\chi$-distribution,
\begin{equation}
\delta f = \sqrt{Y^2}
\label{eq:Delta_f}
\end{equation}
with the centered Gaussian random variable $Y$
with variance $\sigma^2$; $Y \sim {\rm N}(0, \sigma^2)$.
Then a vertex $v^* \in \Omega^{\rm s}_j(t)$ is chosen
so that the pheromone intensity attains the highest value
at that vertex; that is,
$\max_{v \in \Omega^{\rm s}_j(t)} f(v, t)=f(v^*, t)$.
(If the highest intensity of pheromone is taken at
plural vertices, then one of them is randomly chosen.)
At time step $t \to t+1$, the ant hops to 
the chosen vertex,
\begin{equation}
v_j(t) \to v^*.
\label{eq:hopping}
\end{equation}
Notice that the previous position $v_j(t-1)$ is not included
in $\Omega^{\rm s}_j(t)$ in order to
avoid backward walking.

\subsubsection{Random switching at the turning point} 
We introduce a parameter $p \in [0, 1]$.
When $v_j(t)=v_{\rm f}$; that is, just after the ant gets
food, the mode of drifted random walk 
remains to be slow with probability $p$,
and changed to be fast with 
probability $1-p$. Hence, 
when ants are in the homing-period, 
their modes are independently distributed as
\begin{equation}
\sigma_j(t) \equiv
\begin{cases}
{\rm s}, & \mbox{with probability $p$},
\cr
{\rm f}, & \mbox{with probability $1-p$}.
\end{cases}
\label{eq:random_switching}
\end{equation}

\subsubsection{Hopping rules in the homing-period}
\begin{description}
\item{\bf (Slow mode)} \,
In the homing-period, 
if an ant is in the slow mode, $\sigma_j(t) = {\rm s}$, then
the drifted random walk is
similar to that in the exploring-period.
For each ant, $X_j(t)=(v_j(t), {\rm s})$, we define
a set of the nearest-neighboring vertices of $v_j(t)$ as
\begin{align}
\widetilde{\Omega}^{\rm s}_j(t) &:=
\begin{cases}
\{v \in V_L: |v-v_j(t)|=1, v \not= v_j(t-1)\},
\cr
\hskip 6cm
\mbox{if $v_j(t) \in V_L \setminus \{v_{\rm f} \}$},
\cr
\{v \in V_L: |v-v_j(t)|=1\}=\{(L-1, L), (L, L-1) \},
\cr
\hskip 6cm
\mbox{if $v_j(t)=v_{\rm f}$}.
\end{cases}
\label{eq:tilde_Lambda_s}
\end{align}
At each time $t$, the intensities of pheromone
at all vertices in 
$\widetilde{\Omega}^{\rm s}_j(t)$ are listed out 
and add positive
fluctuations to them as \eqref{eq:fluctuation}
with \eqref{eq:Delta_f}. 
Then the vertex $v^* \in \widetilde{\Omega}^{\rm s}_j(t)$ 
is chosen, 
at which the pheromone intensity takes the highest value.
(If the highest intensity of pheromone is taken at
plural vertices, then one of them is randomly chosen.)
At time step $t \to t+1$, the ant hops to the chosen vertex,
\begin{equation}
v_j(t) \to v^*.
\label{eq:hopping2}
\end{equation}
Notice again that the previous position $v_j(t-1)$ is not included
in $\widetilde{\Omega}^{\rm s}_j(t)$ 
in order to forbid backward walking.

\item{\bf (Fast mode)} \,
In the homing-period, 
assume that 
an ant is in the fast mode, $\sigma_j(t) = {\rm f}$.
Then, we define
a subset of the nearest-neighboring vertices of $v_j(t)$ as
\begin{equation}
\widetilde{\Omega}^{\rm f}_j(t) 
:=\{ v \in V_L: |v-v_j(t)|=1, |v| < |v_j(t)|\}.
\label{eq:Lambda_f}
\end{equation}
Then in the time step $t \to t+1$, 
\begin{equation}
v_j(t) \to v 
\quad \mbox{with probability} \quad
\frac{1}{|\widetilde{\Omega}_j^{\rm f}(t)|},
\quad \mbox{if and only if 
$v \in \widetilde{\Omega}^{\rm f}_j(t)$},
\label{eq:fastA1}
\end{equation}
where for a set $S$, $|S|$ denotes the total number 
of element of $S$. 
Notice that since we have considered the system in
a subset $V_L$ of the square lattice, 
$|\widetilde{\Omega}^{\rm f}_j(t)| \in \{1, 2\}$. 
If $x=0$ or $y=0$ in $v_j(t)=(x,y)$, 
$|\widetilde{\Omega}^{\rm f}_j(t)|=1$, and hence, 
the hopping of the fast mode
is deterministic. 
\end{description}

\subsubsection{Initialization and update rule}
\begin{description}
\item{(i)} \,
We start from the initial configuration such that
all $N$ particles are in the slow mode and at the origin
$v_{\rm n}=(0,0)$,
\begin{equation}
X_j(j-1)=(v_{\rm n}, {\rm s}),
\quad j=1, \dots, N.
\label{eq:initial1}
\end{equation}
From $v_{\rm n}$, $N$ particles start  
successively at $t=0,1,\dots, N-1$. 

\item{(ii)} \,
When $v_j(t)=v_{\rm n}$, $t >0$;
that is, an ant comes back to the nest, 
the model becomes to be slow,
$X_j(t)=(v_{\rm n}, {\rm s})$,  
and immediately restarts the exploring walk, 

\item{(iii)} \,
We perform hopping of particles
sequentially according to the numbering of 
particles $j=1, \dots, N$. 
The update of the pheromone field is done at each time step. 
\end{description}


\subsection{Parameter setting}
\label{sec:settings}

We set the parameters concerning 
the time-dependent pheromone field as 
\begin{equation}
f_0=100,  
\quad f_1=5, \quad \delta t=2, \quad f_2=0.1, \quad
\sigma^2=2. 
\label{eq:setting1}
\end{equation}
We assume that 
the density of particles is fixed to be
\begin{equation}
\rho:=\frac{N}{(L+1)^2}=\frac{1}{8}=0.125. 
\label{eq:setting1b}
\end{equation}
We have performed the numerical simulation
of the present stochastic model on the
square lattices with sizes $L=20$, 
$40$, and $60$. 
The time duration of simulation were
$T=1 \times 10^4$,
$2 \times 10^4$, 
$3 \times 10^4$, and $5 \times 10^4$ steps.
In the present paper, 
we will report the dependence of 
the simulation results on the
switching parameter $p$ which is changed in 
the interval $[0,1]$. 
We have confirmed that the numerical results given below
are not changed qualitatively by changing the 
parameter setting \eqref{eq:setting1} 
and \eqref{eq:setting1b}. 

\SSC
{Extinction Transitions of Pheromone-Road 
and Critical Phenomena}
\label{sec:annihilation}
\subsection{Long-term behavior of
survival probability of pheromone-road}
\label{sec:survival}
\begin{figure}[ht]
\begin{center}
\includegraphics[width=0.7\textwidth]{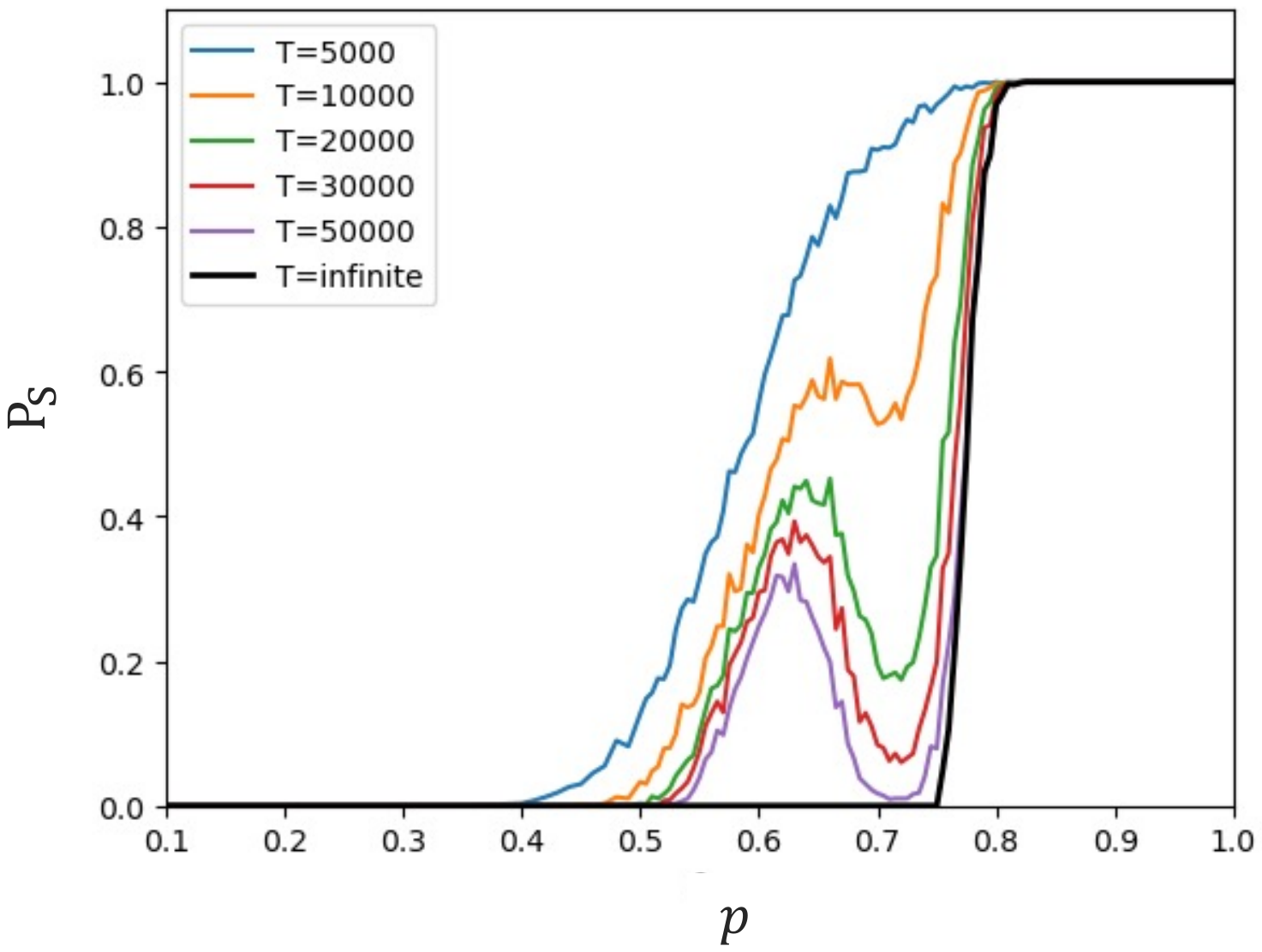}
\end{center}
\caption{Survival probabilities $\PS$ of pheromone-road
are numerically calculated for 
$T=0.5, 1, 2, 3$, and $5 \times 10^4$
in the system of size $L=20$.
The extrapolated values for $T \to \infty$
by \eqref{eq:T_scaling} with \eqref{eq:p^*} 
are also plotted, which are zero
for $p \leq p_{\rm c} = 0.750$
and has positive values for $p > p_{\rm c}$
exhibiting a continuous phase transition from 
the extinction phase to the survival phase of
pheromone-road.
}
\label{fig:order_parameter}
\end{figure}
\begin{figure}[ht]
\begin{center}
\includegraphics[width=0.5\textwidth]{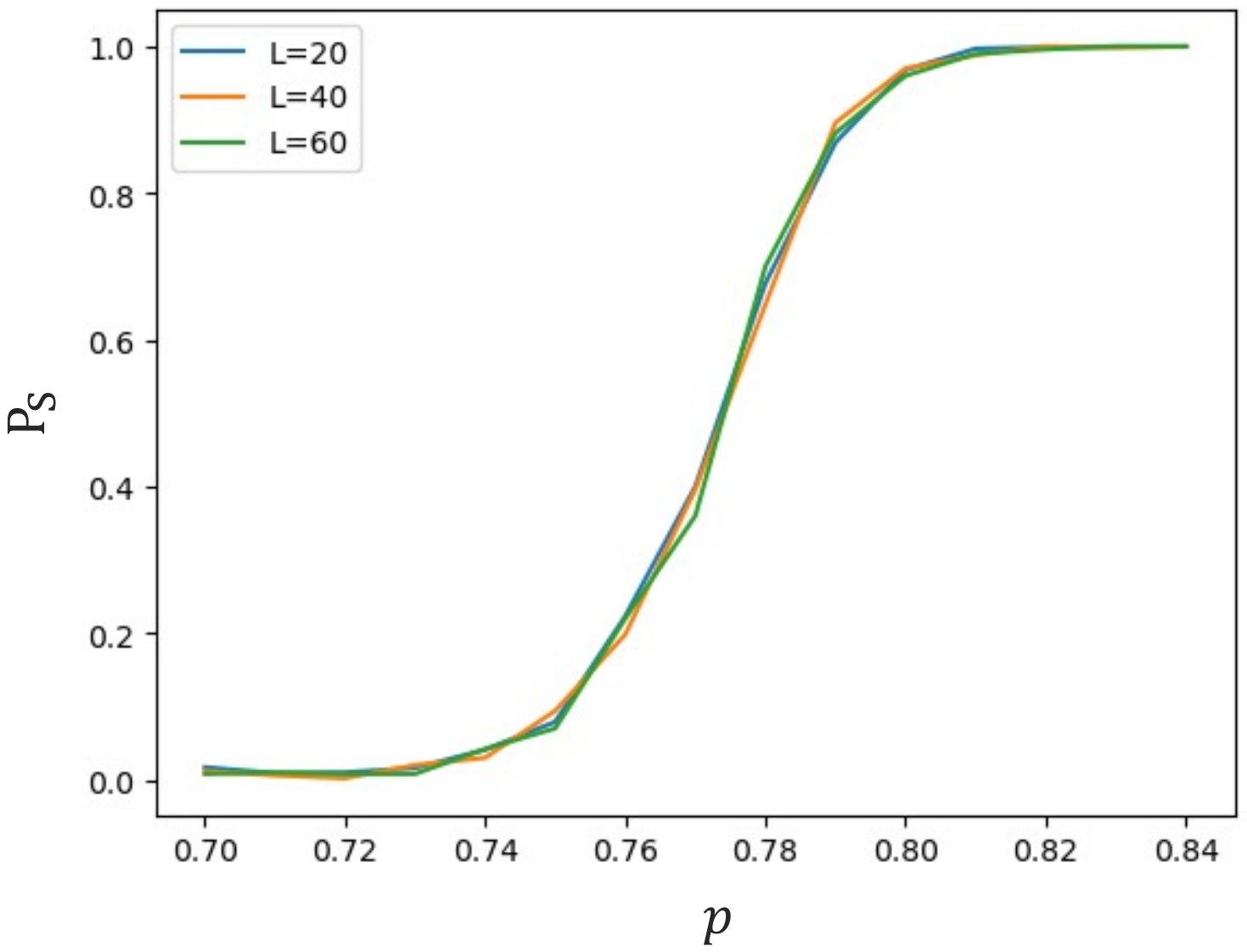}
\end{center}
\caption{Numerically obtained $\PS(p, T, L)$ are
plotted for $L=20$, 40, and 60 with 
fixed value of $T=5 \times 10^4$
for $0.70 \leq p \leq 0.84$.
The dependence of $\PS(p, T, L)$ on 
the system size $L$ seems to be negligible in 
the present model.
}
\label{fig:L_independence}
\end{figure}

We introduce a smallness parameter $\varepsilon \in (0, 1)$
and a time period $\tau > 1$.
At each time step in the numerical simulation,
we observe the value of pheromone field at each vertex
in the $\Gamma$-region $\Gamma_L$ given by
\eqref{eq:Gamma1}.
If 
\begin{equation}
\frac{f(v, t)}{f_0} < \varepsilon 
\quad \mbox{at least one $v \in \Gamma_L$},
\label{eq:disconnect}
\end{equation}
then we say that the pheromone-road is
\textit{disconnected}.
Moreover, if the pheromone-road has been
disconnected during $\tau$ successive steps,
then we say that the pheromone-road \textit{extincts}.
Otherwise, we regard that the pheromone-road survives.

Set a long-term period $T$, and we performed
$M_0$ independent numerical simulations of the model
and counted the number $S(T)$ of events such that
the pheromone-road survives longer than $T$
to obtain the ratio $S(T)/M_0$.
We repeat this procedure $M_1$ times and calculate the
average the ratios to define
the \textit{survival probability} of the pheromone-road
longer than $T$,
\begin{equation}
\PS(p, T, L)
=\left\bra \frac{S(T)}{M_0} \right\ket,
\label{eq:PS}
\end{equation}
which also depends on the switching parameter $p$
and the lattice size $L$.

For each $L$, we set 
\begin{equation}
\varepsilon=1/6, \quad
\tau=5, \quad
M_0=100, \quad M_1=20.
\label{eq:parameterB}
\end{equation}
The $p$-dependence of 
the numerically obtained $\PS(p, T, L)$
are shown for different values of 
$T=5 \times 10^3$, $1 \times 10^4$, $2 \times 10^4$, 
$3 \times 10^4$, and $5 \times 10^4$
in Fig.~\ref{fig:order_parameter},
where $L=20$.
By definition, the survival probability
$\PS(p, T, L)$ should be zero for small vales of $p$
and will approach to 1 as $p \to 1$.
For the intermediate values of $p$, however, 
we see a rather complicated behavior of
as a function of $p$ such that
a local maximum appears around $p=0.6$
for finite values of $T$. 
We will discuss the physical meaning of 
local maximum of $\PS(p, T. L)$
observed around $p=0.6$ in the next section.
On the other hand, the system size $L$ dependence
of $\PS(p, T, L)$ seems to be very small
as demonstrated by Fig.~\ref{fig:L_independence}.
Hence, we neglect the $L$-dependence and
regard that $\PS(p, T, L) \simeq \PS(p, T)$.
We see that independence of
the rescaling \eqref{eq:scaling1} on the size $x$
implies $\nu_{\perp}=0$.

By definition, 
the survival probability $\PS(p, T)$ decreases
monotonically as the time period $T$ increases
and such monotonicity with respect $T$
is clearly observed in Fig.~\ref{fig:order_parameter}.
Numerical analysis shows that the $T$ dependence
is well described by the following power laws,
\begin{equation}
\PS(p, T, L) \simeq \PS(p, T=\infty, L) +
\begin{cases}
c_1 T^{-\eta}, & \, \mbox{for $p < p^*$},
\cr
c_2 T^{-1}, & \, \mbox{for $p \geq p^*$},
\end{cases}
\quad \mbox{as $T \to \infty$},
\label{eq:T_scaling}
\end{equation}
with 
\begin{equation}
p^* \fallingdotseq 0.67, 
\quad \eta \fallingdotseq 0.2,
\label{eq:p^*}
\end{equation}
where $c_1$ and $c_2$ are positive constants.
We evaluated the limit value
$\PS(p, T=\infty, L)$ at each $p \in [0, 1]$
and the results are shown in Fig.~\ref{fig:order_parameter}.
This figure clearly shows that
the critical value of switching parameter is
evaluated as 
\begin{equation}
p_{\rm c} \fallingdotseq 0.750,
\label{eq:pcB}
\end{equation}
and $\PS$ plays a role of the \textit{order parameter}
for the phase transition,
\begin{align}
p < p_{\rm c} \quad \iff \quad
& \mbox{extinction phase of path},
\nonumber\\
p=p_{\rm c} \quad \iff \quad
& \mbox{critical state of path annihilation},
\nonumber\\
p > p_{\rm c} \quad \iff \quad
& \mbox{survival phase of path}.
\label{eq:phase_transition}
\end{align}

\subsection{Finite-size scaling and critical exponents}
\label{sec:finite_size}
\begin{figure}[ht]
\begin{center}
\includegraphics[width=0.5\textwidth]{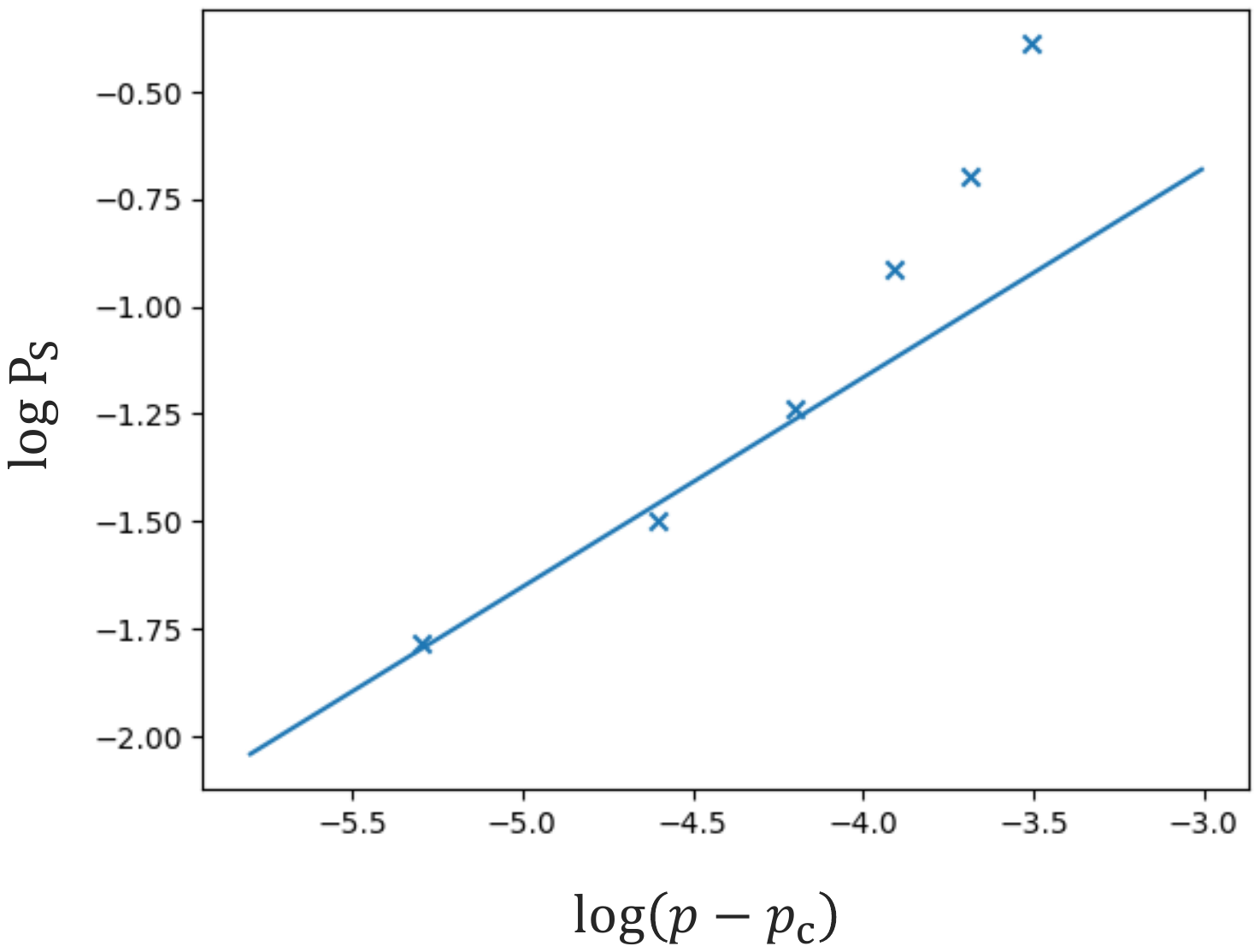}
\end{center}
\caption{Logarithms of $\PS(p, T=\infty)$
are plotted versus $\log(p-p_{\rm c})$
with $p_{\rm c}=0.750$.
The power law \eqref{eq:scaling3a} is confirmed 
with $\beta \fallingdotseq 0.487$.}
\label{fig:beta}
\end{figure}
\begin{figure}[ht]
\begin{center}
\includegraphics[width=0.5\textwidth]{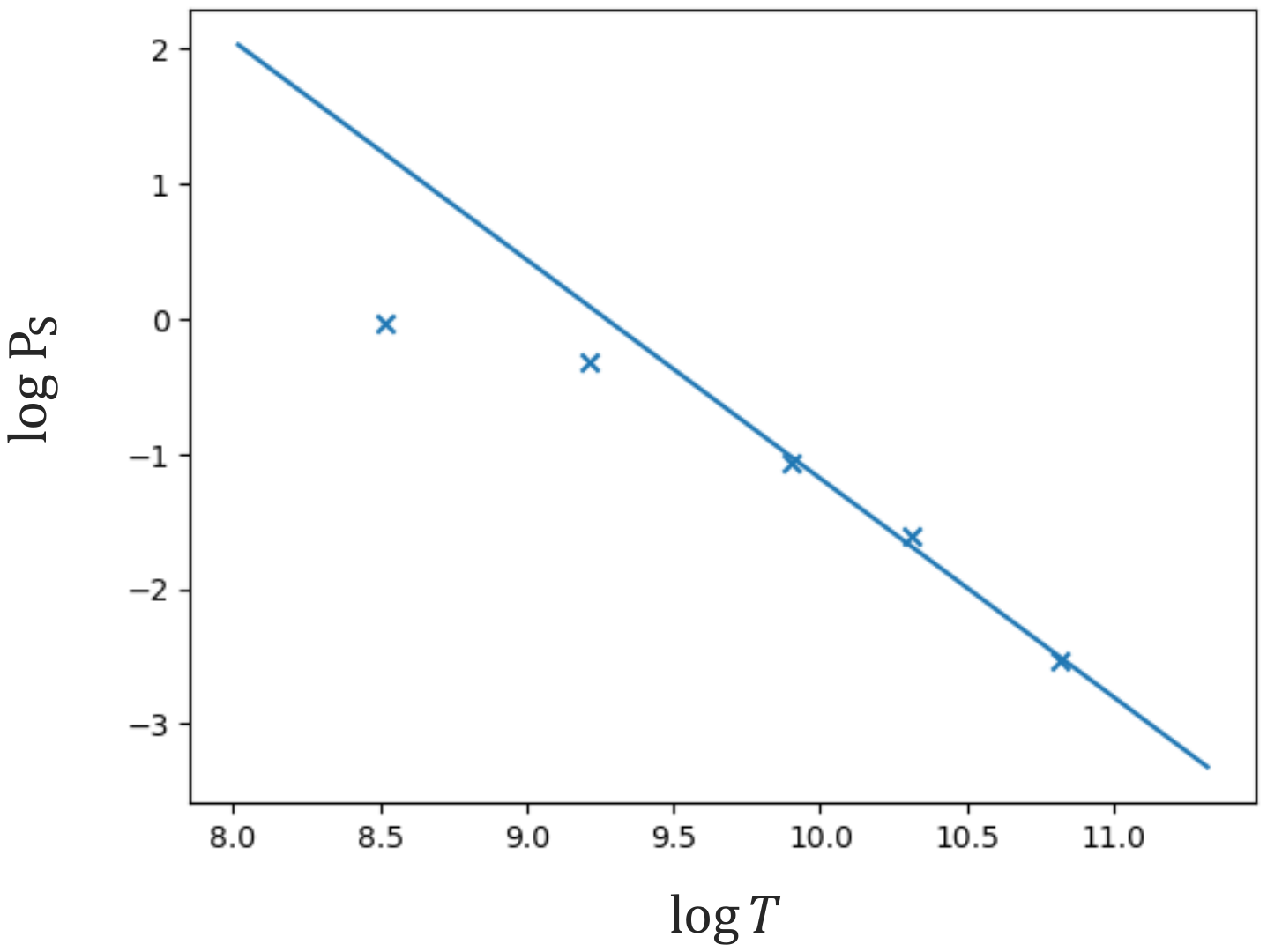}
\end{center}
\caption{Logarithmics of $\PS$
at $p=p_{\rm c}=0.750$ are plotted versus $\log T$. 
The power-law decay \eqref{eq:scaling3b} is confirmed 
with $\beta/\nu_{\parallel} \fallingdotseq 1.62$.}
\label{fig:nu_para}
\end{figure}

As discussed in Section~\ref{sec:introduction}, 
the invariance of the present system under the
scale change \eqref{eq:scaling1} implies that
the survival probability follows the following
scaling law with the positive critical exponents
$\beta$ and $\nu_{\parallel}$, 
\begin{equation}
\PS(p, T)
=T^{-\beta/\nu_{\parallel}}
F(T^{1/\nu_{\parallel}}(p-p_{\rm c})),
\label{eq:scaling2}
\end{equation}
if $0 \leq p-p_{\rm c} \ll 1$ and $T \gg 1$.
Here $F(x)$ is a \textit{scaling function}
and we assume the following asymptotics,
\begin{align}
&\lim_{x \to +\infty} F(x) \sim x^{\beta},
\label{eq:scaling2a}
\\
&\lim_{x \to +0} F(x) \sim 1.
\label{eq:scaling2c}
\end{align}
Here $x \to +0$ means
that $x$ decreases from positive values to zero,
and $\lim f \sim y$ means
$\log f/ \log g \to 1$ in the limit.

First we consider the case that 
$p$ is strictly larger than $p_{\rm c}$ and
$T \to \infty$; that is the system is in the
survival phase of pheromone-road.
In this situation, 
$x=T^{1/\nu_{\parallel}}(p-p_{\rm c}) \to \infty$.
We apply \eqref{eq:scaling2a} to
\eqref{eq:scaling2} and obtain the behavior
of $\PS$ in the vicinity of $p_{\rm c}$
in the survival phase,
\begin{align}
\PS(p) &=
\lim_{T \to \infty}
\PS(p, T)
\sim T^{-\beta/\nu_{\parallel}} 
(T^{1/\nu_{\parallel}}(p-p_{\rm c}))^{\beta}
\nonumber\\
&\sim (p-p_{\rm c})^{\beta},
\quad p \gtrsim p_{\rm c}.
\label{eq:scaling3a}
\end{align}
Figure~\ref{fig:beta} shows the log-log plots of
$\PS(p, T=\infty)$ 
obtained by \eqref{eq:T_scaling} with \eqref{eq:p^*} versus 
$p-p_{\rm c}$ with $p_{\rm c}=0.750$.
The slope gives the estimation of the 
order-parameter critical exponent as
\begin{equation}
\beta \fallingdotseq 0.49.
\label{eq:beta}
\end{equation}

Next we consider the case that $p=p_{\rm c}$,
and then $T \to \infty$.
In this situation, $T^{1/\nu_{\parallel}}(p-p_{\rm c}) \to 0$. 
We apply \eqref{eq:scaling2c} to \eqref{eq:scaling2}
and obtain the power-law decay 
with respect to the time-period $T$ at the critical state,
\begin{align}
\PS(p_{\rm c}, T)
&:= \lim_{p \to +p_{\rm c}}
\PS(p, T)
\sim T^{-\beta/\nu_{\parallel}},
\quad \mbox{as $T \to \infty$}.
\label{eq:scaling3b}
\end{align}
Figure~\ref{fig:nu_para} 
shows the log-log plot of the values of
$\PS$ at $p=p_{\rm c}=0.750$
for the logarithms of $T$ and $L$, respectively.
The power-law decay \eqref{eq:scaling3b} 
is confirmed and
the exponents are evaluated as
\begin{equation}
\beta/\nu_{\parallel} \fallingdotseq 1.6.
\label{eq:exponents2}
\end{equation}
Combination of this value with \eqref{eq:beta}
gives $\nu_{\parallel} \fallingdotseq 0.30$
in \eqref{eq:exponents}.

\SSC
{Creation of New Pheromone-Road and 
Metastability of Old Road}
\label{sec:creation}
\begin{figure}[ht]
\begin{center}
\includegraphics[width=0.6\textwidth]{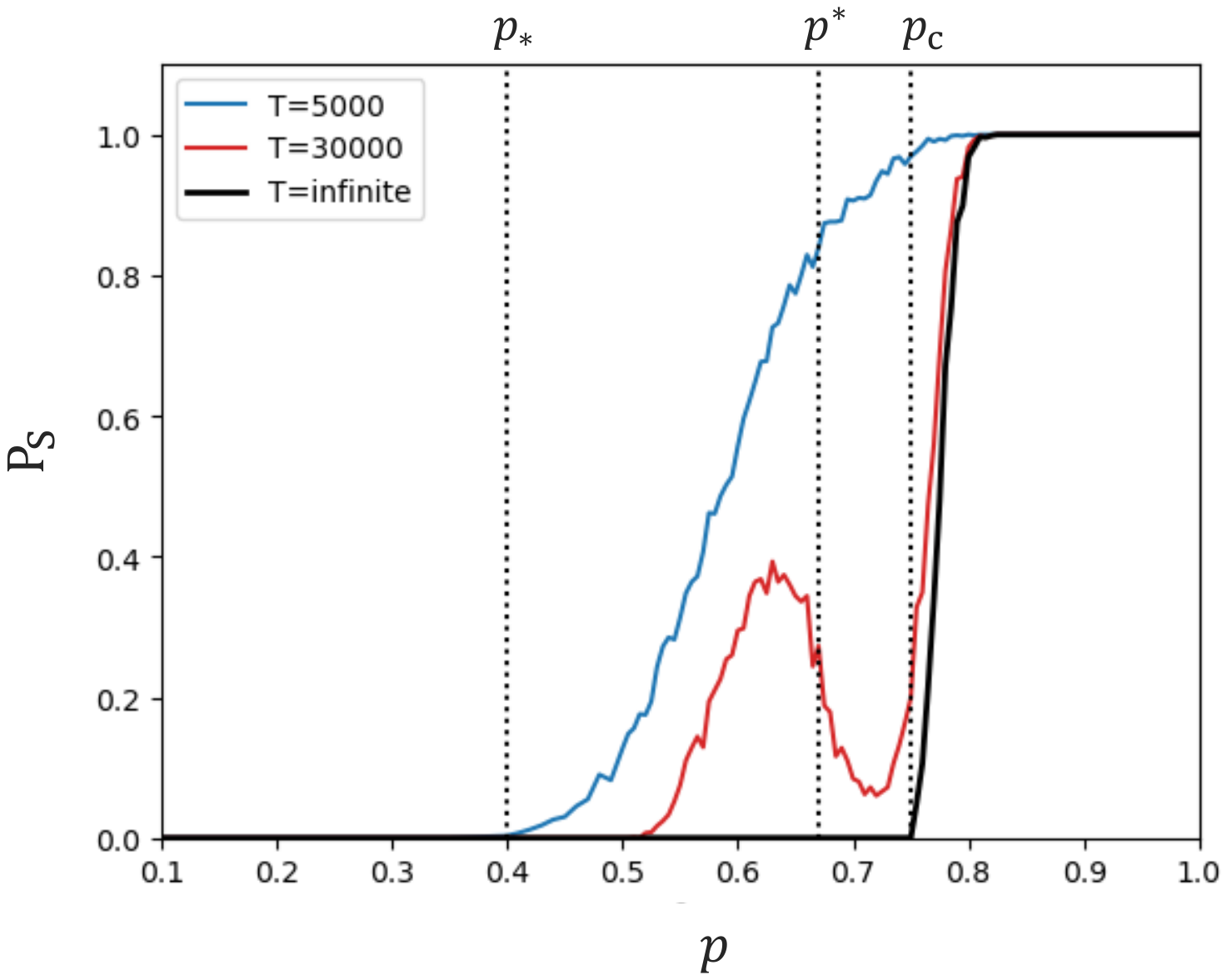}
\end{center}
\caption{
Dependence of observation time $T$ of $\PS(p, T)$ is
classified into four types depending on
the four parameter regimes,
(i) $p_{\rm c} < p \leq 1$,
(ii) $p^* \leq p \leq p_{\rm c}$, 
(iii) $p_* < p < p^*$,
and
(iv) $0 \leq p \leq p_*$.
}
\label{fig:4_regime}
\end{figure}

\begin{figure}[ht]
\begin{center}
\includegraphics[width=0.8\textwidth]{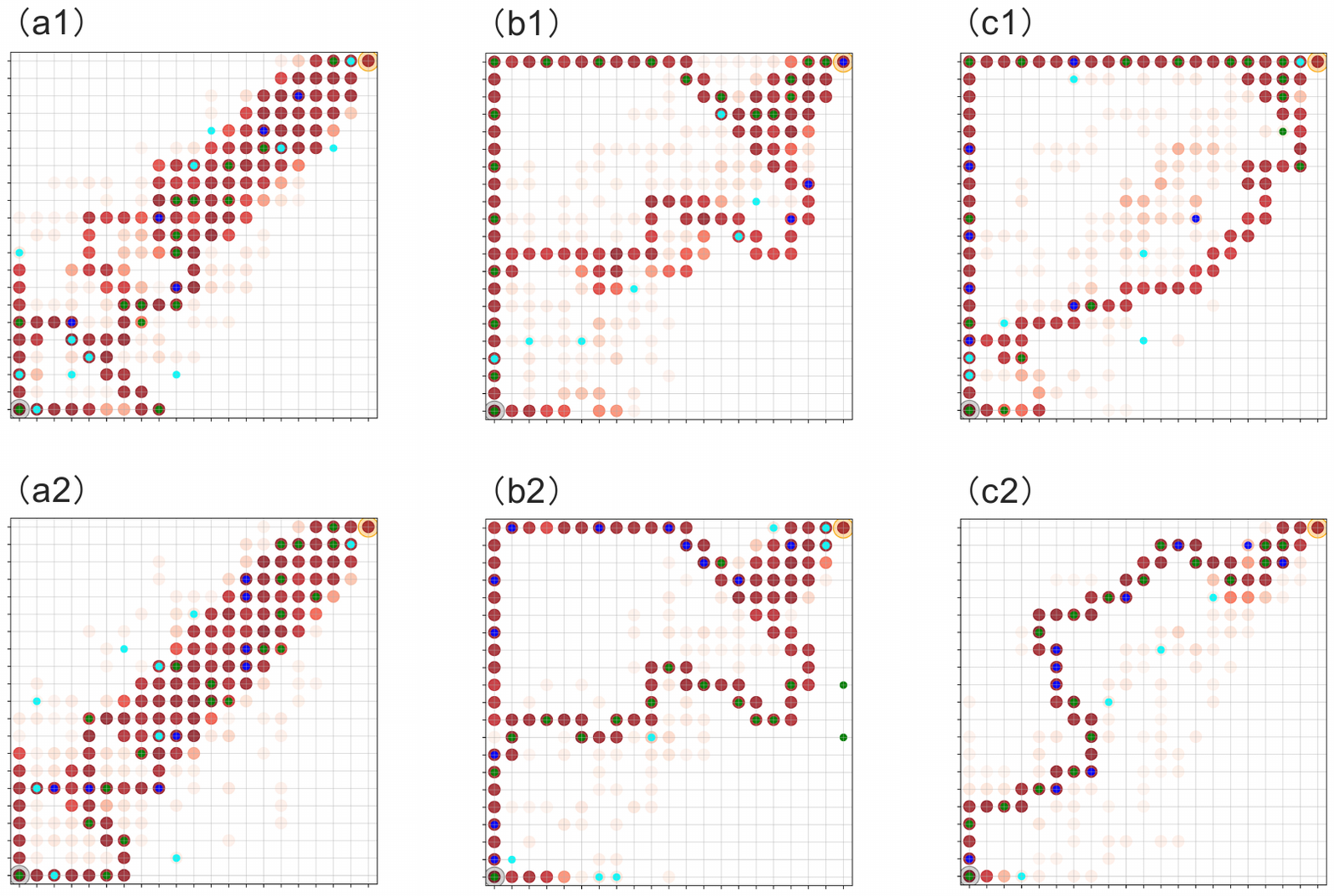}
\end{center}
\caption{
For three values of the parameter, 
(a) $p=0.3$, (b) 0.6, and (c) 0.7,
typical samples of processes are
selected and snapshots of pheromone field 
are shown at a relatively short time
$t=0.5 \times 10^4$ labeled by (1) and at a longer time 
$t=3 \times 10^4$ labeled by (2).
At each figure, the vertices with higher values of
$f(v,t)$ are indicated by darker red.
}
\label{fig:pattern}
\end{figure}

As shown by Fig.~\ref{fig:order_parameter},
the survival probability of the originally prepared
pheromone-road $\PS(p, T)$ exhibits non-monotonic 
behavior with respect to $p$ for finite time 
period $T < \infty$.
This phenomenon will be explained by the
creation of a new pheromone-road consisting of
shorter paths connecting the food source 
and the nest diagonally in the square lattice $V_L$. 
In addition to the threshold value
$p^*=0.67$ introduced in \eqref{eq:p^*}
and the critical value $p_{\rm c}=0.75$ 
for the phase transition associated with
the annihilation of pheromone-road 
\eqref{eq:phase_transition}, 
we introduce another threshold values
\begin{equation}
p_* = 0.40.
\label{eq:p_*}
\end{equation}
As indicated in Fig.~\ref{fig:4_regime}, 
then, the four regimes of parameter $p$ are
considered.

\begin{description}
\item[(i) $p_{\rm c} < p \leq 1$:] \,
In this parameter regime, 
the initially prepared pheromone-road 
survives with probability $\PS(p)$ 
even in the long-term limit $T \to \infty$.
The system is in the survival phase of the
old pheromone-road.
As $p$ decreases from 1 to $p_{\rm c}$,
the homing random walkers in the fast mode
(representing pioneering ants using visual cues) 
increases, who construct a new road which connect
the food source and the nest by direct paths.
The probability to establish the new road in
$T \to \infty$ shall be given by $1-\PS(p)$.
In this sense, the old and new pheromone-road
coexist and the domination of roads is switching
from the old one to the new one as $p$ decreases
in this regime.

\item[(ii) $p^* \leq p \leq p_{\rm c}$:] \,
Since the old pheromone-road was prepared
at the beginning, $\PS(p, T)$ retains positive value
if the observation time $T$ is short, 
but the value is reduced to zero quickly as $T$ increases 
as shown by the blue curve for $T=0.5 \times 10^4$
and red curves for $T=3 \times 10^4$ 
in Fig.~\ref{fig:4_regime}, respectively.
For $p=0.7$ in this parameter regime, 
Fig.~\ref{fig:pattern} (c1) shows a snapshot
of a sample of pheromone filed at time $t=0.5 \times 10^4$,
where the vertices with higher values of
$f(v,t)$ are indicated by darker red.
In such a short time period, the old and new
pheromone roads can coexist.
But as shown by Fig.~\ref{fig:pattern} (c2),
the old road extincts and only the new road 
survives after a long-term $t=3 \times 10^4$.

\item[(iii) $p_* < p < p^*$:] \,
In this parameter regime, approximately
half of homing random walkers is
in the fast mode. They tend to walk directly
from the food source to the nest, but 
the tendency to follow such direct paths
is relatively small compared to
the regime (ii). It means that the new road
will not be reinforced sufficiently by followers.
As a result, the survival time of
the old road becomes longer 
compared to the regime (ii).
This is the reason why we see
local maximums of $\PS(p, T)$ in this parameter
regime (iii) in Fig.~\ref{fig:order_parameter}, 
if we observe the system for finite time-period.
For $p=0.6$, 
Figs.~\ref{fig:pattern} (b1) and (b2) show snapshots
in a sample process of 
pheromone filed at time $t=0.5 \times 10^4$
and $t=3 \times 10^4$, respectively.
Many parts of old road seem to remain for
a long time-period. 
In other wards, in this intermediate parameter
regime, the old road is in a 
\textit{metastable state} and the relaxation time
to be in the extinction phase of the old road
becomes large.

\item[(iv) $0 \leq p \leq p_*$:] \,
In this regime with low values of $p$,
the old pheromone road extinct quickly
as shown by Figs.~\ref{fig:pattern} (a1) and (a2)
for $p=0.3$. Since $p$ is small, many walkers
go directly from the food source to the nest,
but their paths are not reinforced by others.
As a result, the new roads in Figs.~\ref{fig:pattern} (a1) 
and (a2)
are broad and vague compared with
the new road realized in 
Figs.~\ref{fig:pattern} (c2) 
for the parameter regime (ii). 
\end{description}

\SSC
{Concluding Remarks and Future Problems}
\label{sec:future}

In the present paper, we proposed 
a discrete-time stochastic model on a square lattice
to simulate the foraging walks of ants between
the nest and the food source,
which is a modification of the model
introduced by our previous paper \cite{EMTKN24}.
There are two levels of stochastic variables
developing in time $t$. 
One of them is a set of $N$ pairs of
$v_j(t) \in V_L$ and the binary variables
$\sigma_j(t) \in \{{\rm s}, {\rm f}\}$, 
$j=1, 2, \dots, N$, representing the location of 
the walkers (ants), 
which are in one of the two modes, 
slow ({\rm s}: pheromone-mediated walk) or 
fast ({\rm f}: visual-cues-mediated walk).
In the homing walk, 
each ant secretes pheromone
to indicate their paths for allowing latecomers
to trail the paths.
The amount of pheromone is recorded as
variables $f(v, t)$ at each vertex $v \in V_L$,
which represent the pheromone field 
at each time $t$, and the random walkers 
are positively biased by $f(v, t)$, if
the particles are in the slow mode,
$\sigma_j(t)={\rm s}$.
In the previous paper \cite{EMTKN24}, 
the global changes of the trajectories of
walkers were studied
depending on a switching parameter,
while in the present paper
the global changes of the spatial distribution
of the pheromone field $f(v, t)$ are studied
depending on the switching parameter $p$. 
Since the time-evolution of $f(v, t)$ is slower
than the walks of particles, here we made
long-term simulations over
$T = 10^{4}$ steps and evaluated 
the $T \to \infty$ limits by performing
$T^{-\eta}$-fitting.

In the present work, we first focused on the
extinction of initially prepared pheromone-road, 
but the above consideration
suggests that this problem is connected with 
the problem how a new road is created. 
The evaluated critical exponents \eqref{eq:exponents}
are quite different from the values in
the DP universality class \cite{CD98,Hin00,MD99}.
The critical phenomena associated with 
\eqref{eq:exponents} should be clarified.

As shown by Fig.~\ref{fig:pattern},
the development of the new road from the food source
to the nest is highly anisotropic in space; that is, two kinds of
correlation lengths shall be considered as
$\xi_{\rm diag}$ along the diagonal direction
and $\xi_{\rm ortho}$ along its orthogonal direction.
As a matter of course, we need a time-direction
to describe such non-equilibrium phenomena.
In addition to the correlation length in time direction
$\xi_{\parallel}$,  
the above two kinds of spatial correlation length
$\xi_{\rm diag}$ and $\xi_{\rm ortho}$ shall be
regarded as the two types of
correlation lengths perpendicular to $\xi_{\parallel}$
and will be denoted by $\xi_{\perp}^{(\ell)}$ with
$\ell=1,2$. 
Study of such 
\textit{multiply-directed percolation problem}
is a challenging future problem
from the view point of non-equilibrium statistical
mechanics. 

We hope that the present study of pheromone-roads
simulated by the mathematical models for 
foraging ants
will give insights into 
general problems to characterize 
creation and extinction phenomena 
of paths and roads in the real world.

\vskip 1cm
\noindent{\bf Acknowledgements} \,
MK was supported by JSPS KAKENHI Grant Numbers 
JP19K03674, 
JP21H04432,
JP22H05105,
JP23K25774, 
and
JP24K06888.
HN was supported by JSPS KAKENHI Grant Numbers 
JP20H01871, 
JP21H05293,
and
JP21H05297.



\begin{thebibliography}{99} 
\bibitem{CD98}
B. Chopard, M. Droz, 
Cellular Automata Modeling of Physical Systems
(Cambridge University Press, 
Cambridge, 1998).

\bibitem{dH22a}
F. den Hollander, S. Nandan, 
Spatially inhomogeneous populations with seed-banks:
I. Duality, existence and clustering, 
J. Theor. Probab. 
\textbf{35} (2022) 1792--1841.

\bibitem{dH22b}
F. den Hollander, S. Nandan, 
Spatially inhomogeneous populations with seed-banks:
II. Clustering regime, 
Stochastic Process. Appl.
\textbf{150} (2022) 116--146. 

\bibitem{EMTKN24}
A. Ezoe, S. Morimoto, Y. Tanaka, M. Katori, H. Nishimori,
Switching particle systems for foraging ants 
showing phase transitions in path selections,
Physica A
\textbf{643} (2024) 129798 (13 pages). 

\bibitem{Flo22}
S. Floreani, C. Giardin\`a, F. den Hollander, S. Nandan, F. Redig, 
Switching interacting particle systems:
Scaling limits, uphill diffusion and boundary layer, 
J. Stat. Phys. 
\textbf{186} (2022) 33 (45 pages). 

\bibitem{Gre22}
A. Greven, F. den Hollander, M. Oomen, 
Spatial populations with seed-bank:
Well-posedness, duality and equilibrium, 
Electon, J. Probab.
\textbf{27} (18) (2022) 1--88.

\bibitem{Gre23}
A. Greven, F. den Hollander, M. Oomen,
Spatial populations with seed-bank:
Renormalisation on the hierarchical group, 
to appear in Memoirs of AMS, 
arXiv:math.PR/2110.02714.

\bibitem{HSIMT18}
T. Hashimoto, K. Sato, G. Ichinose, R. Miyazaki,
K. Tainaka,
Clustering effect on the dynamics in a spatial
rock-paper-scissors system,
J. Phys. Soc. Jpn.
\textbf{87}, 014801 (2018).

\bibitem{Hin00}
H. Hinrichsen,
Non-equilibrium critical phenomena and phase transitions
into absorbing states,
Adv. Phys. 
\textbf{49} (7) (2000) 815--958.

\bibitem{HW90}
B. H\"{o}lldobler, E. O. Wilson, 
The Ants, 
(Harvard University Press, Cambridge, MA, 1990).

\bibitem{LSAJAH16}
G. Lemoult, L. Shi, K. Avila, S. V. Jalikop, 
M. Avila, B. Hof,
Directed percolation phase transition to sustained
turbulence in Couette flow, 
Nat. Phys.
\textbf{12} (2016) 254--258.

\bibitem{Len12}
J. T. Lennon, F. den Hollander, M. Wilke-Berenguer, J. Blath, 
Principle of seed banks and the emergence
of complexity from dormancy, 
Nat. Commun. 
\textbf{12} (2021) 4807 (16 pages). 

\bibitem{MD99}
J. Marro, R. Dickman, 
Nonequilibrium Phase Transitions in
Lattice Models, 
(Cambridge University Press, 
Cambridge, 1999).

\bibitem{MOSS92}
H. Matsuda, N. Ogita, A. Sasaki, K. Sat\=o,
Statistical mechanics of population:
the lattice Lotka--Volterra model,
Prog. Theor. Phys.
\textbf{88}, 1035--1049 (1992).

\bibitem{MKN25}
S. Morimoto, M. Katori, H. Nishimori,
Interacting particle systems modeling self-propelled motions, 
{\sf arXiv:2502.08543[cond-mat,stat-mech]}.

\bibitem{NMI97}
M. Nakanura, H. Matsuda, Y. Iwasa,
The evolution of cooperation in a lattice-structured 
population,
J. Theor. Biol. 
\textbf{184}, 65--81 (1997).

\bibitem{NSN17}
H. Nishimori, N. J. Suematsu, S. Nakata,
Collective behavior of camphor floats migrating
on the water surface,
J. Phys. Soc. Jpn.
\textbf{86}, 101012 (2017).

\bibitem{Nis15}
Y. Ogihara, O. Yamanaka, T. Akino, S. Izumi, A. Awazu, H. Nishimori, 
Switching of primarily relied information by ants:
A combinatorial study of experiment and modeling, 
in: (eds.) T. Ohira, T. Uzawa (Eds.), 
Mathematical Approaches to Biological Systems
--Networks, Oscillations, and Collective Motions,
Springer-Verlag, 2015, 
pp.119--137.

\bibitem{ST16}
M. Sano, K. Tamai,
A universal transition to turbulence in
channel flow
Nat. Phys.
\textbf{12} (2016) 249--253.

\bibitem{Shi13}
S. Shinoda, 
Analysis of the influence of visual cues for
foraging ants (in Japanese).
Graduate Thesis,
Department of Mathematics, Hiroshima University,
Hiroshima (2013) [in Japanese].

\bibitem{Tai88}
K. Tainaka,
Lattice model for the Lotka--Volterra system,
J. Phys. Soc. Jpn. 
\textbf{57}, 2588--2590 (1988).

\bibitem{TKCS07}
K. A. Takeuchi, M. Kuroda, H. Chat\'e, M. Sano,
Directed percolation criticality in turbulent liquid crystals,
Phys. Rev. Lett.
\textbf{99} (2007) 2345003 (4 pages).

\bibitem{TKCS09}
K. A. Takeuchi, M. Kuroda, H. Chat\'e, M. Sano,
Experimental realization of directed percolation criticality in turbulent liquid crystals, 
Phys. Rev. E
\textbf{80} (2009) 051116 (12 pages).

\bibitem{VZ12}
T. Vicsek, A. Zafeiris, 
Collective motion.
Phys. Rep. 
\textbf{517} (2012) 71--140. 

\end{thebibliography}
\end{document}